# MACHINE-LEARNING PREDICTION OF THE COMPUTED BAND GAPS OF DOUBLE PEROVSKITE MATERIALS

Junfei Zhang[1], Yueqi Li[2], and Xinbo Zhou[3]

[1] School of Computing and Information Systems, The University of Melbourne, Melbourne, Victoria, Australia
[2] College of Physical Science and Technology, Xiamen University, Xiamen, Fujian, China
[3] Faculty of Information Technology, Beijing University of Technology, Beijing, China

## ABSTRACT

*Prediction of the electronic structure of functional materials is essential for the engineering of new devices. Conventional electronic structure prediction methods based on density functional theory (DFT) suffer from not only high computational cost, but also limited accuracy arising from the approximations of the exchange-correlation functional. Surrogate methods based on machine learning have garnered much attention as a viable alternative to bypass these limitations, especially in the prediction of solid-state band gaps, which motivated this research study. Herein, we construct a random forest regression model for band gaps of double perovskite materials, using a dataset of 1306 band gaps computed with the GLLBSC (Gritsenko, van Leeuwen, van Lenthe, and Baerends solid correlation) functional. Among the 20 physical features employed, we find that the bulk modulus, superconductivity temperature, and cation electronegativity exhibit the highest importance scores, consistent with the physics of the underlying electronic structure. Using the top 10 features, a model accuracy of 85.6% with a root mean square error of 0.64 eV is obtained, comparable to previous studies. Our results are significant in the sense that they attest to the potential of machine learning regressions for the rapid screening of promising candidate functional materials.*

## 1. INTRODUCTION

In quantum mechanics, the energy of bound electrons becomes quantized [1], and electrons at the ground state can be excited to higher energy levels by absorbing photons with the corresponding wavelengths. In solid structures, the superposed electronic states form continuous energy bands. In insulators and semiconductors, the band gap is the energy gap across the valence and conduction band where electrons are forbidden to occupy. The magnitude of the band gap plays an important role in many functional materials, such as transistors, photovoltaics, light-emitting diodes, and sensors [2]. For instance, optoelectronic materials are generally wide-band gap semiconductors, while thermoelectric materials are narrow-band gap semiconductors [3]. Hence, accurate and efficient prediction of band gaps of solid materials is crucial for the design and engineering of new devices.

One of the most widely used electronic structure methods for evaluating band gaps is density functional theory (DFT) [4]. In the Kohn-Sham formalism [5], the multielectron wavefunction is replaced by fictitious noninteracting states that give rise to the true electron density [6], which enables the iterative solution of the single-particle Hamiltonian. However, the exchange-correlation energy, which contains all the quantum mechanical interactions of the electrons, does not have an exact expression in terms of the electron density and as such requires an approximation, such as the local density approximation (LDA) [7] or the generalized gradient approximation (GGA) [8]. Such approximations have limited accuracy, most notably the underestimation of the band gap of semiconductors and insulators [9]. Various approaches have been proposed to address this limitation, such as the on-site Hubbard $U$ correction [10], hybrid functionals using fractional exact exchange [11], and quasiparticle methods such as the GW approximation [12]. However, these methods do not always guarantee an accurate description of the system, and they can be much more computationally expensive than conventional DFT [13].

An alternative strategy for band gap prediction is machine learning. For example, a support vector regression model was constructed for inorganic solids using experimentally measured band gaps [14], thereby bypassing the limitations of DFT. Another study trained a kernel ridge regression model [15] using band gaps computed with the GLLBSC (Gritsenko, van Leeuwen, van Len the, and Baerends solid correlation) functional [16], which demonstrated reasonable agreement with experimental values. These studies attest to the potential of machine learning methods, provided that robust datasets are available for training [17]. The importance of band gap prediction of functional materials and the above-mentioned limitation of DFT serves as the motivation for this research study, which attests to the potential of machine learning regression for band gap prediction.

We employ a dataset of GLLBSC-computed band gaps of 1306 double perovskites in this study. Double perovskites ($AA'BB'X_6$) have double the unit cell of single perovskites ($ABX_3$) with chemically distinct $A/A'$ and $B/B'$ sites [18]. A variety of physical and chemical properties can be engineered by doping the cations with species of different valence states or radii [19]. Due to their stable crystal structure, unique electromagnetic properties, and high catalytic activities, these compounds have much potential as functional materials for environmental protection [20], the chemical industry [21], photovoltaics [22], and catalysis [23]. In this regard, optimization and engineering in the above-mentioned fields require a proper description of the underlying electronic structure of double perovskites [24], which attests to the significance of choosing the band gaps of double perovskites as our dataset.

Previous studies have shown that random forest regression is well-suited to capturing nonlinearity, as seen across the band gap and the extracted physical features such as the highest occupied energy level [25]. As such, we construct a random forest regression model for predicting the band gap of double perovskite compounds, building upon a previous kernel ridge regression study [15]. We find that the bulk modulus, superconductivity temperature, and cation electro negativity exhibit the highest importance scores among the 20 physical descriptors employed, consistent with the physics of the underlying electronic structure. A model accuracy of 85.6% with a root mean square error of 0.64 eV is obtained using the top 10 features, comparable to previous studies [1].

The succeeding part of the paper is structured as follows: The literature review is given in section 2; the research methodology is presented in section 3; section 4 presents the results and discussion, including an evaluation of the performance of our model as well as our limitations; finally, section 5 gives the concluding remarks of this work.

## 2. LITERATURE REVIEW

This research study focuses on the prediction of the band gaps of double perovskite materials using machine learning, as a surrogate method for the conventional prediction yielded by the DFT. The limitation of the DFT, notably the lack of expression of the exchange-correlation energy, and the potential of machine learning in solving the issue have urged computer scientists to try various machine learning models for band gap prediction. This section will review recently proposed machine learning models for band gap prediction.

### 2.1 Tuplewise Graph Neural Networks (TGNN)

Na, G. S. et al. [26] conducted a research study using modified TGNN (Tuplewise Graph Neural Networks) to predict the band gap of a crystalline compound. TGNN is designed to automatically generate crystal representation using crystal structures and to include the crystal-level properties as an input feature. In this study, the prediction of the band gap using TGNN is shown to have higher accuracy than the standard DFT. The results of two out of four datasets that the study employed are of interest in our research: 1345 organic-inorganic perovskite materials of which the targeting band gap is the hybrid screened exchange functional (HSE06) and 2233 materials for solar cells with the targeting band gap as GLLBSC-computed band gap. Using the proposed TGNN model, the experiment of the former dataset achieved an MAE of 0.045 eV and that of the latter dataset achieved an MAE of 0.295 eV.

### 2.2. Alternating Conditional Expectations (ACE)

ACE (Alternating Conditional Expectations) is a machine learning algorithm designed to find the optimal transformation between the two sets of variables, and performs well on small data sets; its advantage is that the results are represented in graphic form. The limitation of ACE is that if the dependence of the response variable on the predictors is slightly different than the transformation that the algorithm estimated, the analytic formulas are very difficult to discover. Gladkikh, V. et al. [27] conducted a study exploring the mappings between the band gap and the properties of the constituent elements using ACE. The study employs a dataset containing a large number of single perovskite materials ($ABX_3$). The best result achieved using ACE has an RMSE of 0.836 eV and an MAE of 0.602 eV.

### 2.3. Kernel Ridge Regression (KRR)

Regonia, P.R. et al. [28] trained a KRR (Kernel Ridge Regression) model for the prediction of the optical band gap of zinc oxide ($ZnO$). Kernel ridge regression is a variant of ridge regression that is suitable for small datasets and is usually used for the prediction of the band gap of organic crystal structures. The model is trained using two empirical features: the experimental time and temperature conditions during $ZnO$ fabrication. Quadratic features are generated to increase the model's complexity and prevent the dataset's underfitting. The result presents an RMSE of 0.0849 eV.

## 3. METHODS

### 3.1. Random Forest Regression

Random forest regression is a regression method that utilizes multiple decision trees, which are constructed by a simple supervised algorithm consisting of a series of if-then-else statements. The randomness is manifested through random sampling of data subsets or random selection of features. Multiple uncorrelated decision trees construct a random forest, where all trees are granted free growth without any pruning. The random forest algorithm can be employed for both classification and regression. For classification, the result is the outcome with the highest turnout among all trees; for regression, the forest takes the average of all trees. The steps to generate a random forest are as follows (**Fig. 1** illustrates a flow chart of the algorithm):

1. From a sample with capacity $N$, conduct bootstrap sampling $K$ times. The resulting $K$ samples are used as the node samples of decision trees.
2. Choose a constant $m$ smaller than the dataset feature number $M$.
3. When splitting each decision tree, select $m$ features from the original $M$ features, choosing one feature as the splitting feature of the node. The Gini index is used to calculate the information gain and determine the splitting.
4. Repeat step 2 and step 3 for each node until no splitting can occur, when the next feature is used by the parent node in the last splitting. The tree is always left unpruned to ensure free growth.
5. Repeat steps 1-4 to generate a random forest.

A random forest can manage data with a high dimension of features without performing dimension reduction or feature selection. This is beneficial for the dataset of this study, which involves multiple atomic descriptors of double perovskites. The mutual effects of different features and their significance are also quantified. Although random forest regression is computationally efficient and accurate when using a large number of generated trees, the risk of overfitting still exists for data with a large noise. We perform random forest regression as implemented in *scikit-learn*, using the *double_perovskites_gap* dataset available in the *matminer* package [29]. Comparing previous literature, which is normally trained using 5 to 10 atomic features [30], our result is unique in the sense that we use a total of 20 atomic features to achieve a more comprehensive result, of which the dimension is then reduced to 10 features. The selected important features are also consistent with the underlying physics, making the results more credible.

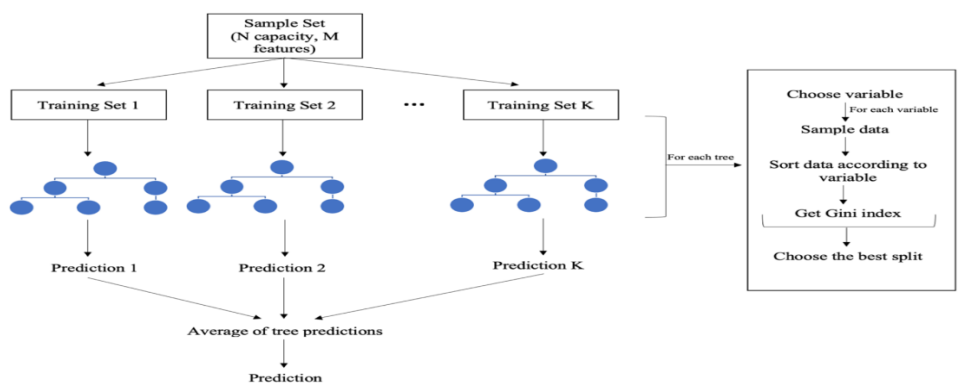

Figure 1. Flow chart of random forest regression

## 3.2. Features

20 atomic features are obtained from the *periodic_table* and *composition* modules of the *Pymatgen* [31] (Python Materials Genomics) package:

　　　Average electronegativity
　　　Average cation electronegativity

Average atomic radius
Average van der walls radius
Average Mendeleev number
Average electrical resistivity
Average molar volume
Average thermal conductivity
Average boiling point
Average melting point
Average critical temperature
Average superconduction temperature
Average bulk modulus
Average youngs modulus
Average Brinell hardness
Average rigidity modulus
Average mineral hardness
Average Vickers hardness
Average density of solid phase
Average first ionization energy

The dataset is first converted into a data frame, which is then processed by applying the chemical composition of each compound to corresponding classes and functions in the *Pymatgen* package to obtain the 20 features. Compositional averages are taken for atomic features of a given compound, whereas molecular features are used directly. Missing values are not counted in the calculation of the average.

## 4. RESULTS AND DISCUSSION

### 4.1. Model Selection

Random forest regression has two parameters to be optimized: the number of estimators (*n_estimator*) referring to the number of trees to be built before taking the maximum voting or averages of predictions; and the random seed (*random_state*) for the random generator. Both the accuracy and the computational cost of the model increase with the number of estimators [32]. The cost scales as $O(n_{\text{tree}} * m_{\text{try}} * n \log(n))$, where $n_{\text{tree}}$ is the number of estimators, $m_{\text{try}}$ is the number of variables to sample at each node, and $n$ is the number of records [33]. As such, an optimal number of estimators is needed to ensure a satisfactory model performance.

As shown in **Fig. 2**, the model accuracy reaches a maximum at around 700 estimators and decreases afterward, which is attributed to overfitting. As such, the *n_estimator* is set to 700. On the other hand, the random seed determines the random sampling for the train-test split and may subtly affect the accuracy due to the randomization of the training pipeline. An optimal *random_state* value of 14 is selected.

The corresponding parity plot of the model prediction is shown in **Fig. 3**. Using a test/training ratio of 0.25 and all 20 physical descriptors, the model accuracy is 85.1% with a mean absolute error (MAE) of 0.47 eV, a root mean squared error (RMSE) of 0.62 eV, which is comparable to the RMSE value of 0.5 eV reported in a previous kernel ridge regression study of the same dataset.

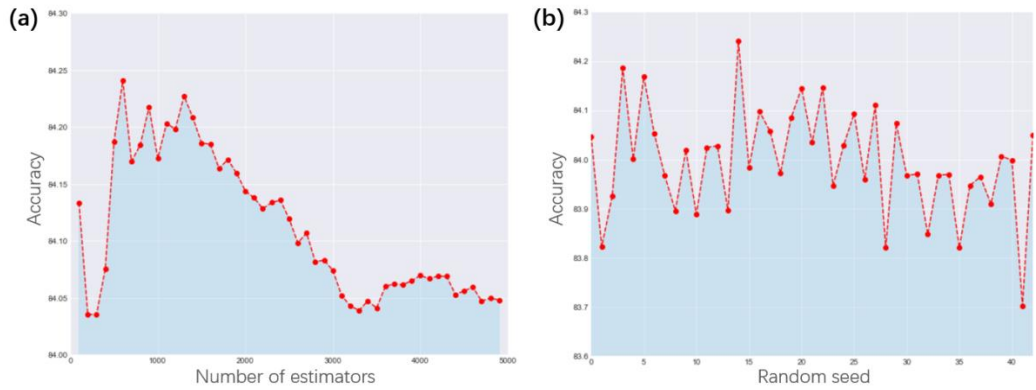

Figure 2. The accuracy of the random forest regression model as a function of (**a**) the number of estimators and (**b**) the random seed, using all 20 physical descriptors.

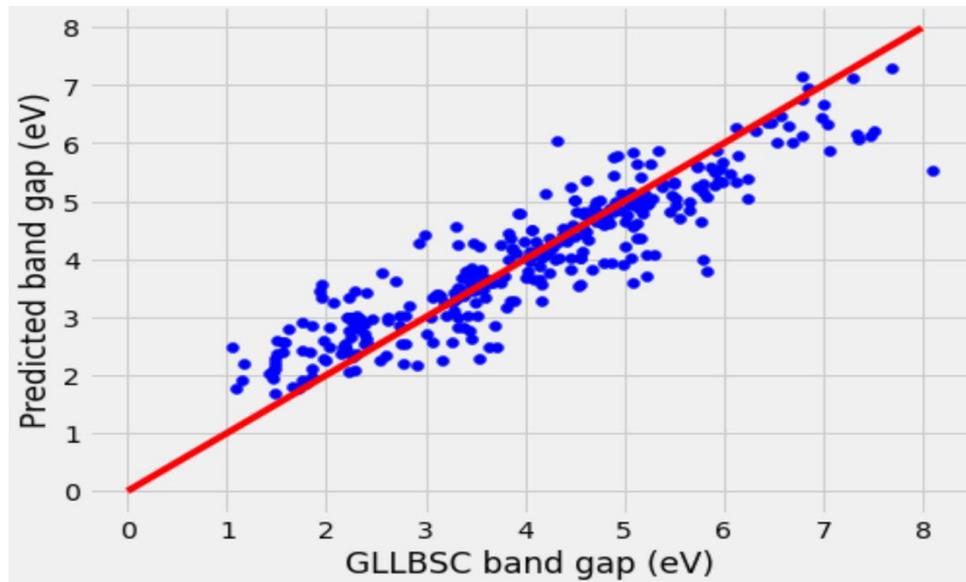

Figure 3. Parity plot of the predicted vs. GLLBSC-computed band gaps, obtained using all 20 physical descriptors and a test/training ratio of 25/75. The parity line is shown in red.

## 4.2. Feature Selection

The feature importance plot is shown in **Fig. 4**. The top three features with the highest importance scores are average bulk modulus, superconductivity temperature, and cation electronegativity:

1)      Bulk modulus quantifies the elastic property of a solid or fluid under pressure, specifically its resistance to compression [34]. Microscopically, bulk modulus depends on the compressibility of atoms, which affects the extent of the overlap of valence atomic orbitals, and therefore the band gap of the material [35].

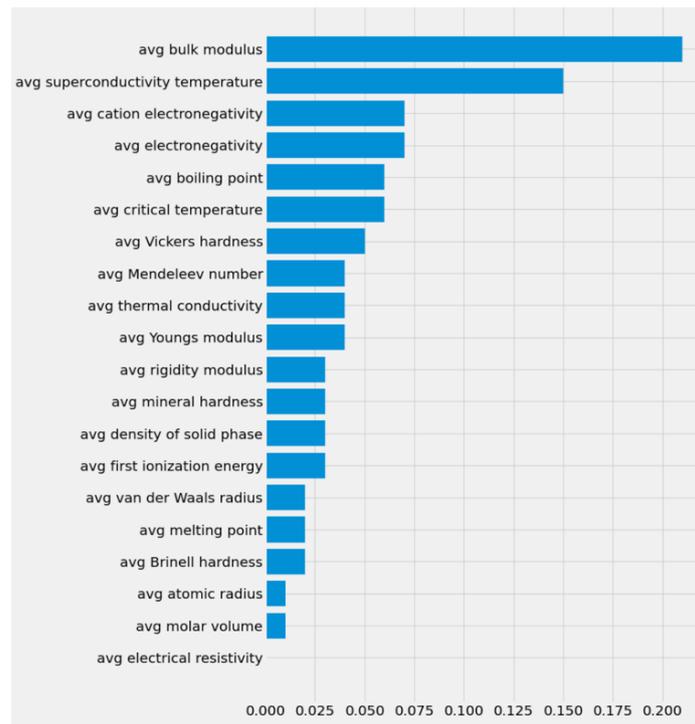

Figure 4. Feature importance of all 20 physical descriptors, obtained from a test/training ratio of 25/75.

2)      Superconductivity is the state of matter with no electrical resistance and magnetic penetrability [36]. Given that the magnitude of the band gap determines the electrical conductivity, a material with a relatively small band gap is expected to more easily achieve a superconducting state [37].

3)      Electronegativity quantifies the ability of an atom to attract an electron pair in a chemical bond [38]. The cation electronegativity here refers to the electronegativity difference between the oxygen anions and the metal cations. A larger elemental electronegativity difference leads to a larger degree of electron localization around the more electronegative element, which makes it harder for electrons to leap to the conduction band [39].

The low importance scores of some features, such as average electrical resistivity and molar volume, indicate that the dataset contains a large amount of noise, which necessitates feature selection. **Table 1** summarizes the model performance using different numbers of top features. The performance remains optimal up to the top 10 features, which yields an accuracy of 85.6% with an RMSE of 0.64 eV. Given the marginal difference in accuracy using 20, 15, and 10 top features, the remainder of the study employs the top 10 features only.

Table 1. The model performance obtained using different numbers of features with the highest feature importance scores (MAE = mean absolute error; RMSE = root mean squared error; NRMSE = normalized RMSE).

| Number of top features | 20 | 15 | 10 | 5 | 3 | 1 |
|---|---|---|---|---|---|---|
| Accuracy (%) | 85.1 | 85.5 | 85.6 | 82.3 | 82.4 | 65.2 |
| MAE (eV) | 0.47 | 0.46 | 0.46 | 0.56 | 0.57 | 1.12 |
| RMSE (eV) | 0.62 | 0.62 | 0.64 | 0.79 | 0.81 | 1.43 |
| NRMSE | 0.08 | 0.07 | 0.08 | 0.10 | 0.10 | 0.17 |

The corresponding importance scores and parity plots are shown in **Figs. 5** & **6**, respectively. The model constructed using the top 10 features exhibits the least deviation of the data points from the parity line. Moreover, the models overall tend to show a larger underestimation for larger band gap values, which can potentially be attributed to the limited accuracy of the GLLBSC functional itself [40].

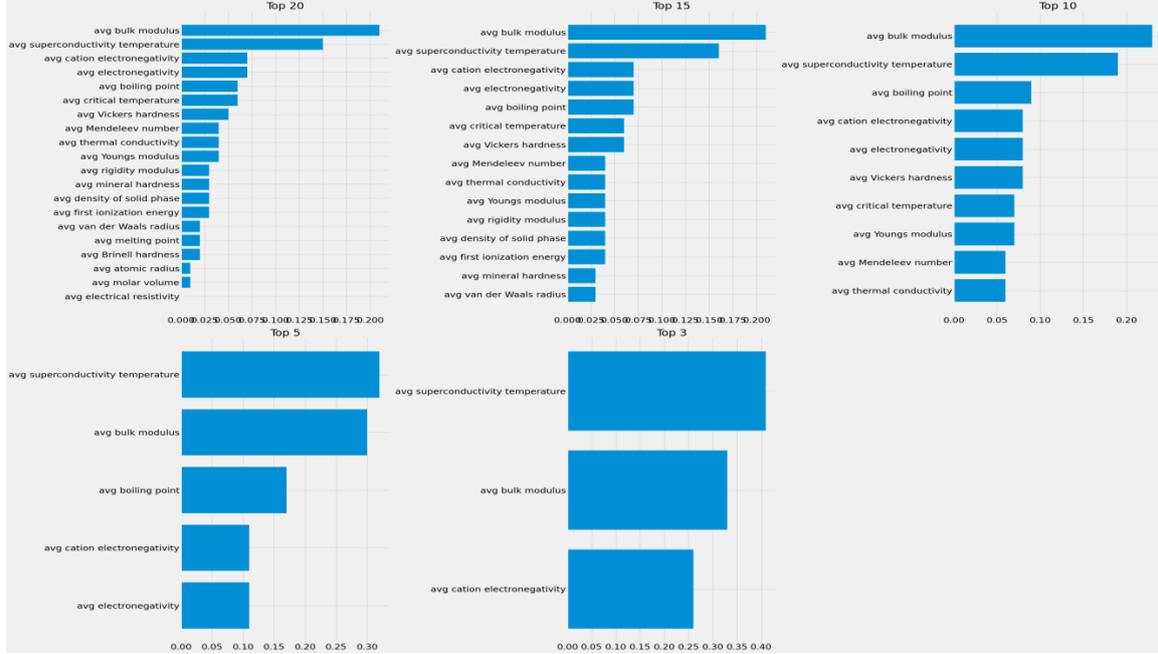

Figure 5. Feature importance scores for models constructed using a number of features with the highest importance scores.

## 4.3. Testing and Training Set partition

**Table 2** summarizes the model performance as a function of the different test-to-training set partitions, ranging from 10/90 to 75/25. As expected, the test set accuracy decreases with the number of training set data points. The corresponding parity plots in **Fig. 7** also demonstrate a larger extent of deviation from the parity line as the proportion of the training set decreases. Based on these results, we validate that the test/training ratio of 25/75 is sufficient in providing satisfactory accuracy (85.6%) and reasonable RMSE (0.64 eV).

Table 2. Model performance obtained with different test-to-training set partitions.

| Test/training set ratio | 10/90 | 20/80 | 25/75 | 40/60 | 50/50 | 75/25 |
|---|---|---|---|---|---|---|
| Number of test set data points | 131 | 262 | 327 | 523 | 653 | 980 |
| Number of training set data points | 1175 | 1044 | 979 | 783 | 653 | 326 |
| Test set accuracy (%) | 87.9 | 86.8 | 85.6 | 82.6 | 82.5 | 76.2 |
| MAE (eV) | 0.41 | 0.45 | 0.46 | 0.5 | 0.53 | 0.67 |
| RMSE (eV) | 0.57 | 0.63 | 0.64 | 0.7 | 0.74 | 0.88 |
| NRMSE | 0.07 | 0.08 | 0.08 | 0.08 | 0.09 | 0.11 |

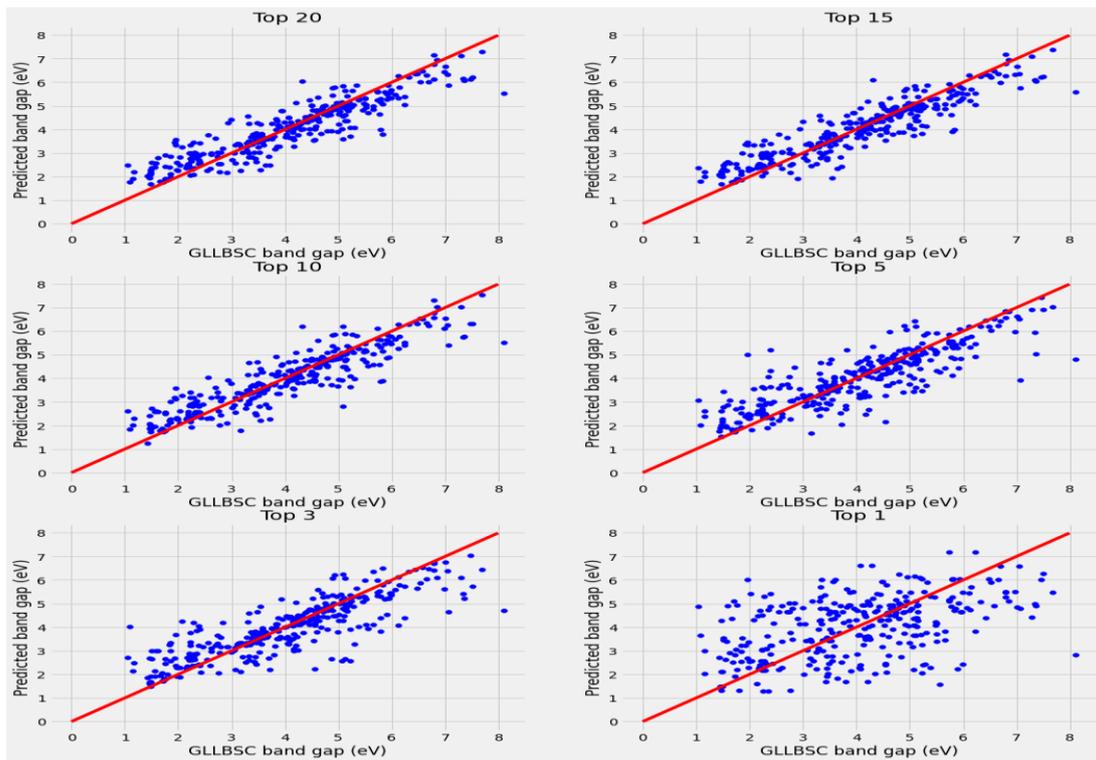

Figure 6. Parity plots of the predicted vs. GLLBSC-computed band gaps obtained using different numbers of features with the highest importance scores. The parity line is shown in red.

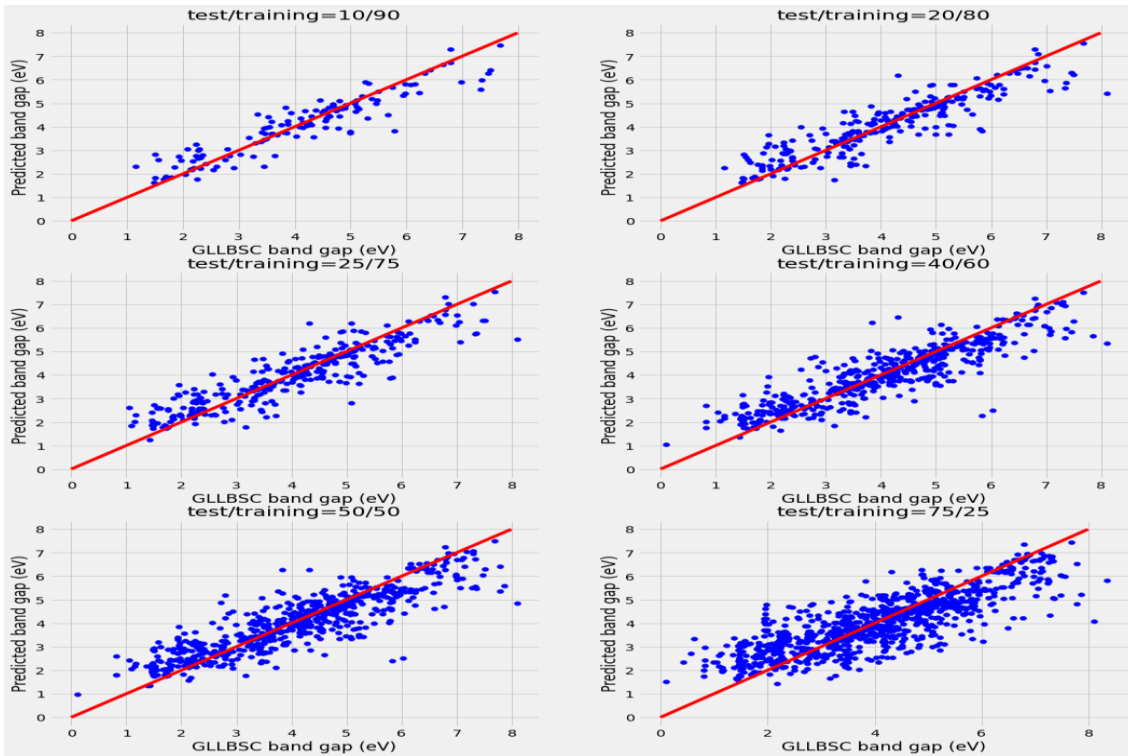

Figure 7. Parity plots of the predicted vs. GLLBSC-computed band gaps obtained using different test-to-training set partitions. The parity line is shown in red.

## 4.4. Model Performances

**Table 3** summarizes the result of previous studies. The best performance yields in KRR by P. R. Regonia et al. [28] with an RMSE of 0.09. Our random forest regression model is comparable to linear regression and XGBoost by G. S. Na et al. [26] and has a lower MAE than ACE and ET by V. Gladkikh et al. [27].

Table 3. Results of the models for band gap prediction (TGNN = tuplewise graph neural networks; XGBoost = extreme gradient boosting; ACE = alternating conditional expectations; ET = extremely randomized trees; KRR = kernel ridge regression; ANN = alternating conditional expectations; GBR = gradient boosting regression).

| Model | Study | Material type | Number of materials | Band gap | Accuracy (%) | MAE (eV) | RMSE (eV) |
|-------|-------|---------------|---------------------|----------|--------------|----------|-----------|
| Random forest | J. Zhang et al. | Double perovskites | 1306 | GLLBSC | 85.6 | 0.46 | 0.64 |
| TGNN | G. S. Na et al. [26] | Materials for solar cells | 2233 | GLLBSC | - | 0.30 | - |
| Linear regression | G. S. Na et al. [26] | Materials for solar cells | 2233 | GLLBSC | - | 0.44 | - |
| XGBoost | G. S. Na et al. [26] | Materials for solar cells | 2233 | GLLBSC | - | 0.44 | - |
| ACE | V. Gladkikh et al. [27] | Single perovskites | - | HSE | - | 0.60 | 0.84 |
| ET | V. Gladkikh et al. [27] | Single perovskites | - | HSE | - | 0.54 | 0.75 |
| KRR | P. R. Regonia et al. [28] | *ZnO* quantum dots | - | Optical band gap | 98.0 | - | 0.09 |
| ANN | P. R. Regonia et al. [28] | *ZnO* quantum dots | - | Optical band gap | 97.8 | - | 0.09 |
| GBR | M. Guo et al. [8] | Binary compounds | 4096 | DFT-calculated band gap | 81.0 | - | 0.26 |

## 4.5. Limitations and Recommendations

This study is limited by the relatively small sample size. We use 1306 data to generate all the results, which may reduce the power of the study and cause a large margin of error. Future research studies can focus on using larger datasets, which we suppose will improve the model fitting. In this study, the missing values are filled by the mean value of that feature. This preprocessing step can be taken more carefully by trying various means to deal with the missing values. Another limitation of the study is that we lack a more interpretable understanding of random forest regression in statistical learning theory. A single decision tree is interpretable because it follows several decision steps, whereas a forest lacks this step-by-step interpretability. Hence, using interpretability tools such as the RF Visualization Toolkits [41] to generate a "Decision Path View" may help to understand the forest. This is essential since the feature's importance is related to the underlying physics.

## 5. CONCLUSION

Despite the widespread use of first-principles methods based on density functional theory (DFT) in materials science, it remains computationally costly and limited in its accuracy due to the approximation of the exchange-correlation functional. In this regard, machine learning presents a viable alternative for the rapid prediction of materials' electronic properties while retaining reasonable fidelity to DFT. This study has implemented random forest regression for the prediction of the band gap of double perovskite compounds employing a dataset of 1306 GLLBSC-computed band gaps. Among the 20 physical descriptors, average bulk modulus, superconductivity temperature, and cation electronegativity exhibited the highest importance scores, which provide a physically interpretable description in terms of the underlying electronic structure. Optimal model performance is obtained with the top 10 features and a test/training partition of 25/75, yielding a model accuracy of 85.6% and RMSE of 0.64 eV comparable to previous studies. Our results highlight the potential of machine learning regression for rapid and physically interpretable prediction of the electronic properties of functional materials.

### ACKNOWLEDGMENTS

This work was supported by Touch Education Technology Inc. We acknowledge scientific and editorial support from the Project Lead, J. S. Lim of Harvard University; technical support from the Project Support C. Zhang; and administrative support from C. Ding of Touch Education Technology Inc.

This work was led by J.Z. with support from Y.L. and X.Z. J.Z. performed machine learning, literature review, and drafted the manuscript. Y.L. performed parameter optimization, visualization, and literature review. X.Z. assisted with literature review and writing.

## AUTHORS AND CO-AUTHORS

**Junfei Zhang** is the author of this paper. She is currently pursuing a Bachelor of Science degree at The University of Melbourne, Melbourne, Victoria, Australia. She is actively conducting computer science related research studies. Her research interest includes machine learning, computer vision, and quantum algorithms. She is currently studying Computer Science at The University of Copenhagen, Copenhagen, Denmark as an exchange student at the time of publication.

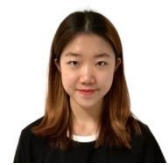

**Yueqi Li** is the co-author of this paper. She is pursuing B.S. Physics in College of Physical Science and Technology, Xiamen University, China. Her main areas of research interest are Biophysics and Machine learning.

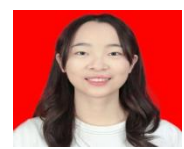

**Xinbo Zhou** is the co-author of this paper. She is a junior student in the Faculty of Information Technology at Beijing University of Technology

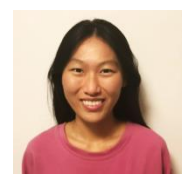